\documentclass[10pt,letterpaper]{article}
\usepackage{amsmath}
\usepackage{graphicx}

\def\fig{Fig.\,}
\def\Figure{Figure\,}
\def\eq{Eq.\,}

\begin{document}

\title{Creation and Measurement of Broadband Squeezed Vacuum from a Ring Optical Parametric Oscillator}

\author{Takahiro Serikawa, Jun-ichi Yoshikawa, Kenzo Makino\\ and Akira Frusawa}

\maketitle

Department of Applied Physics, School of Engineering, The University of Tokyo, 7-3-1 Hongo, Bunkyo-ku, Tokyo, 113-8656, Japan

akiraf@ap.t.u-tokyo.ac.jp

\begin{abstract}
We report a 65\,MHz-bandwidth triangular-shaped optical parametric oscillator (OPO) for squeezed vacuum generation at 860\,nm. The triangle structure of our OPO enables the round-trip length to reach 45\,mm as a ring cavity, which provides a counter circulating optical path available for introducing a probe beam or generating another squeezed vacuum. Hence our OPO is suitable for the applications in high-speed quantum information processing where two or more squeezed vacua form a complicated interferometer, like continuous-variable quantum teleportation. With a homemade, broadband and low-loss homodyne detector, a direct measurement shows $8.4\,\mathrm{dB}$ of squeezing at 3\,MHz and also $2.4$\,dB of squeezing at 100\,MHz.
\end{abstract}

\def\opex{ Opt.\ Express }
\def\boe{ Biomed.\ Opt.\ Express }
\def\ome{ Opt.\ Mater.\ Express }
\def\ao{ Appl.\  Opt.\ }
\def\aop{ Adv.\ Opt.\ Photon.\ }
\def\ap{ Appl.\  Phys.\ }
\def\apa{ Appl.\  Phys.\ A }
\def\apb{ Appl.\  Phys.\ B }
\def\apl{ Appl.\ Phys.\ Lett.\ }
\def\apj{ Astrophys.\ J.\ }
\def\bell{ Bell Syst.\ Tech.\ J.\ }
\def\jqe{ IEEE J.\ Quantum Electron.\ }
\def\assp{ IEEE Trans.\ Acoust.\ Speech Signal Process.\ }
\def\aprop{ IEEE Trans.\ Antennas Propag.\ }
\def\mtt{ IEEE Trans.\ Microwave Theory Tech.\ }
\def\iovs{ Invest.\ Ophthalmol.\ Visual\ Sci.\ }
\def\jcp{ J.\ Chem.\ Phys.\ }
\def\jmo{ J.\ Mod.\ Opt.\ }
\def\jocn{ J.\ Opt.\ Commun.\ Netw.\ }
\def\jon{ J.\ Opt.\ Netw.\ }
\def\josa{ J.\ Opt.\ Soc.\ Am.\ }
\def\josaa{ J.\ Opt.\ Soc.\ Am.\ A }
\def\josab{ J.\ Opt.\ Soc.\ Am.\ B }
\def\jpp{ J.\ Phys.\ }
\def\nat{ Nature }
\def\oc{ Opt.\ Commun.\ }
\def\ol{ Opt.\ Lett.\ }
\def\opn{ Opt.\ Photon.\ News }
\def\pl{ Phys.\ Lett.\ }
\def\pra{ Phys.\ Rev.\ A }
\def\prb{ Phys.\ Rev.\ B }
\def\prc{ Phys.\ Rev.\ C }
\def\prd{ Phys.\ Rev.\ D }
\def\pre{ Phys.\ Rev.\ E }
\def\prl{ Phys.\ Rev.\ Lett.\ }
\def\pr{ Photon.\ Res.\ }
\def\rmp{ Rev.\ Mod.\ Phys.\ }
\def\pspie{ Proc.\ SPIE }
\def\sjqe{ Sov.\ J.\ Quantum Electron.\ }
\def\vr{ Vision Res.\ }
\def\cleo{ {\it Conference on Lasers and Electro-Optics }}
\def\assl{ {\it Advanced Solid State Lasers }}
\def\tops{ Trends in Optics and Photonics }

\section{Introduction}
A squeezed vacuum is a non-classical state of light, which suppresses quantum fluctuation in one quadrature beyond the shot noise level \cite{Walls83}. Taking advantage of the reduced fluctuation, squeezed vacua are expected to contribute to metrological applications like gravitational wave detectors \cite{Kimble01,Goda08,Vahlbruch10} and quantum imaging \cite{Kolobov93,Lu12}. A squeezed vacuum is generated as an even-number photon stream via 2nd or 3rd nonlinear optical effect. Consisting of photon pairs, a squeezed vacuum is combined with a single photon detector to pursue quantum state engineering. Generation of non-Gaussian states of light, such as a Schr\"{o}dinger's kitten state \cite{Ourjoumtsev06,Nielsen06,Wakui07}, is demonstrated in this way. Another point of view is the fact that Einstein-Podolsky-Rosen (EPR) state is easily created by employing two squeezed vacua and a beam splitter \cite{Furusawa98}. Therefore a squeezed vacuum can be considered as a resource of quantum entanglement and is applied in quantum information processing schemes like continuous-variable (CV) quantum teleportation \cite{Furusawa98,Lee11}. Recently, time-domain multiplexing technique exploits this feature and huge scale entanglement is created \cite{Yokoyama13,Yoshikawa16}. In \cite{Yokoyama13,Yoshikawa16}, continuous-wave (CW) squeezed vacua are virtually divided into wavepackets to form a specific entangled state (an extended EPR state) after going through beam splitters followed by delay-lines which rearrange the temporal order of wavepackets. Generation of time-bin qubits is also demonstrated in similar way using a CW squeezed vacuum and a delay-line \cite{Takeda12}. Based on these techniques, further expansion of temporal and also spatial mode numbers of the entangled state will lead to measurement-based one-way quantum computing \cite{Raussendorf01} or CV cluster computing \cite{Gu09}. However, this is not easy due to the instability caused by the long delay-line, which has to contain one virtual wavepacket at a time. To shorten the delay-line and accelerate the processing rate, a broadband squeezed vacuum is required since the temporal length of virtual wavepackets is limited by the bandwidth of CW squeezed vacua \cite{Takei06}.

Quadrature squeezing was first observed in four-wave mixing \cite{Slusher85}, and later, optical parametric process was used \cite{Wu86}. Variety of methods have been investigated to obtain squeezing, for example, a CW pumped optical parametric oscillator (OPO) \cite{Yurke84,Collett84}, a $\chi^{(2)}$ material waveguide \cite{Anderson95,Serkland95,Eto11}, and a $\chi^{(3)}$ process in a optical fiber \cite{Shelby86,Bergman91}. $15$\,dB of squeezing is recently reported \cite{Vahlbruch16} with an OPO-type squeezed vacuum source and till now there is no other way to realize above $6$\,dB of CW vacuum squeezing. An advantage of OPO-type squeezed vacuum sources is the enhancement of the nonlinear effect by longitudinal confinement. With an OPO, the pump-induced optical loss in the nonlinear crystal can be suppressed, since less optical power is required for pumping and even a finite pump power is sufficient to provide any parametric gain. However, the bandwidth of a squeezed vacuum is limited by the cavity structure. Although the bandwidth of nonlinear crystals is over 100\,GHz deriving from the phase-matching condition, we can typically make use of only a narrow bandwidth of a single resonant peak of an OPO cavity, whose linewidth is determined by output coupler transmissivity $T$ and round-trip length $l$. Here we can say that $l$ should be small to make the linewidth broad, whereas $T$ should be kept at certain value to maintain the longitudinal confinement.

An example of a short, broadband OPO is a monolithic-type one \cite{Breitenbach95}, where the two surfaces of a nonlinear crystal works as mirrors and they forms a Fabrry-Perrot cavity. The simple structure of a monolithic OPO enables the round-trip length to be below 10\,mm, resulting the top record of the bandwidth of 2.26\,GHz as a CW squeezed vacuum from an OPO \cite{Ast13}. However, note that a monolithic cavity has only one free parameter, temperature, to be controlled for cavity resonance and phase matching. For the applications with a single OPO like gravitational wave detectors, the laser frequency could be tuned. On the other hand, when we focus on the applications in CV quantum information processing, this solution is not acceptable because two or more squeezed vacua must be in the same frequency to interfere. A semi-monolithic OPO \cite{Laurat05, Zhou15} is an alternative of a monolithic OPO, which has an external mirror to form a Fabrry-Perrot cavity together with the one surface of the nonlinear crystal. A semi-monolithic OPO does not have the problem in resonance-controlling because the cavity length can be tuned independently. Nevertheless, these Fabrry-Perrot type OPOs have another issue in common, that is, a cavity-locking probe beam runs along the squeezed vacuum and gets in the successive optical circuits. If the probe is prepared in a frequency shifted mode or a orthogonal polarization mode, it is possible to eliminate the probe beam from squeezed vacuum. However, extra optical loss will be induced in the picking-up setup especially for frequency-shifted probe. Moreover, for orthogonal polarization probe, detuning may be caused by the birefringence of the nonlinear crystal, which is affected by the fluctuation of temperature or the pumping power. Another available option would be optical chopping of the probe, however, it might limit the time scale of the sequence of the entangled state \cite{Yokoyama13,Yoshikawa16}. A ring cavity OPO, in contrast, possesses an advantage of having two circulating beam paths. One of them can be dedicated to the squeezed vacuum, while the other can be used for introducing the probe beam. Then the probe beam can be spatially separated, even if it is in the same frequency and polarization as the squeezed vacuum. Additionally, a ring OPO has many mirrors which allows cavity-probing beams or phase-reference beams of the squeezed vacuum to pass through. With backward pumping, the counter circulating path affords even one more squeezed vacuum, as long as the cavity-probing problems as in Fabrry-Perrot cavities mentioned above is ignorable. Owing to these experimental flexibility, there are some instances of ring OPOs, often bow-tie type OPOs, utilized in the large-scale quantum optical experiments with two or more OPOs like CV teleportation \cite{Furusawa98,Lee11} or entanglement generation \cite{Yokoyama13,Yoshikawa16}. While $9.0$\,dB of squeezing is reported \cite{Takeno07} with a bow-tie OPO, the bandwidth of a bow-tie OPO is typically below 20\,MHz corresponding to the minimum length of a virtual wavepacket of 30\,m \cite{Yokoyama13} due to the complex structure.

In this paper, we demonstrate a generation of a 65\,MHz bandwidth squeezed vacuum with a triangle-shaped ring OPO, whose round-trip length is 45\,mm. For the specification of the squeezing spectrum, a broadband optical homodyne detection up to 200\,MHz is performed with a newly developed photodetector. The squeezing levels of $8.4$\,dB around DC and $2.5$\,dB at 100\,MHz are observed without any noise correction.

\section{Design of OPO}

\begin{figure}
 \centering
 \includegraphics[width=12cm]{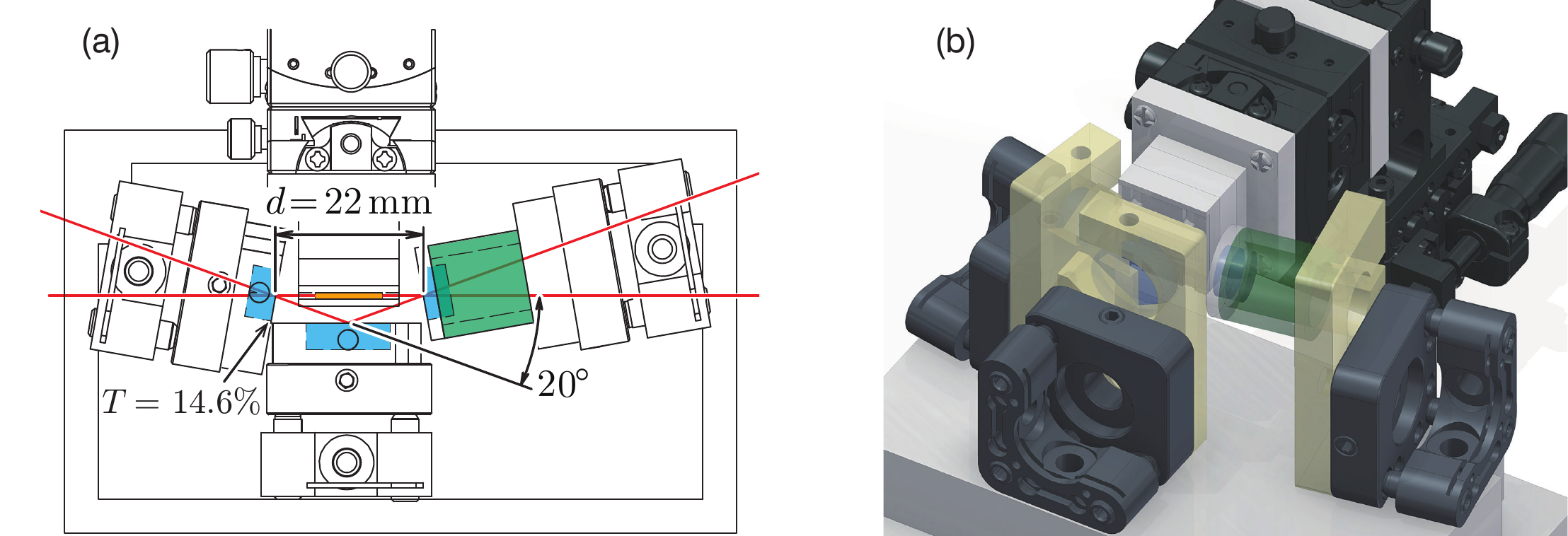}
 \caption{(a) Design of our OPO. Red lines are optical path, blue boxes are mirrors and green box is a piezo actuator. The PPKTP crystal is depicted as a orange box at the long edge of triangle. $d$ = 22\,mm is the distance between spherical mirrors. (b) Schematic picture.} \label{fig:opo}
\end{figure}

Figure \ref{fig:opo} shows the schematic of our OPO. A triangle cavity, whose round-trip length is 45\,mm and corresponding optical path length $l$ is 53\,mm, consists of three mirrors. Two spherical mirrors (LAYERTEC, radius of curvature $R$ = 15.0\,mm, diameter $\phi$ = 6.35\,mm, angle of incidence AOI = $10^\circ$), are placed at the vertexes of the acute angles. One of them is the output-coupler mirror with partial-reflection coating (power transmissivity $T$ = 14.6\% at 860\,nm) and the other is high-reflection (HR) mirror. Linewidth of the cavity is 65\,MHz expressed in $f_\mathrm{HWHM} = cT / 4\pi l$. To make a cavity-length control, the HR spherical mirror is attached to a ring piezo actuator (Pizomechanik, HPSt 150/14-10/12). The remaining mirror is a flat HR mirror (LAYERTEC, $\phi$ = 12.5\,mm, AOI = 70). Two HR mirrors have a small transmissivity $T$ = 300\,ppm, through which control beams can be introduced or picked up. A periodically poled $\mathrm{KTiOPO}_4$ (PPKTP) crystal (Raicol Crystals, 1.0\,mm$\times$1.0\,mm$\times$10.0\,mm) is placed at the long edge of the triangle, affording 10\,mm long interaction length.

\begin{figure}
 \centering
 \includegraphics[width=6.5cm]{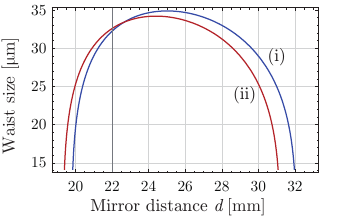}
 \caption{Beam waist size in the PPKTP crystal of the resonant mode of the OPO. (i) Horizontal. (ii) Vertical. $d$ is the distance between two spherical mirrors as shown in \fig\ref{fig:opo}. The gray line represents the actual value of $d$ = 22\,mm. } \label{fig:mode}
\end{figure}

Note that the triangle structure demands relatively large folding angle compared to a bow-tie cavity, which may cause the transverse mode to be elliptic, since the effective curvatures of spherical mirrors are different between horizontal and vertical axes. Additionally, the obliquely placed output-coupler mirror acts as an asymmetric lens when the squeezed beam is extracted. The transverse mode of a squeezed vacuum is highly desired to be circular when we consider applications in which squeezed vacuum interferes with other beams. In order to minimize the distortion of the shape of the output beam, we optimized the geometric parameter of the cavity. \Figure\ref{fig:mode} shows the calculation of the waist size at the center of the long edge of the triangle. The horizontal and vertical diameters have different dependences on the distance $d$ of the two spherical mirrors and meet at $d =$ 22.7\,mm. Considering the lens effect of the output-coupler, we set $d$ at $22.0$\,mm expecting 99.9\% modematch to the optimal circular $\mathrm{TEM}_{00}$ beam.

\section{Experiment}
\begin{figure}
 \centering
 \includegraphics[width=13.5cm]{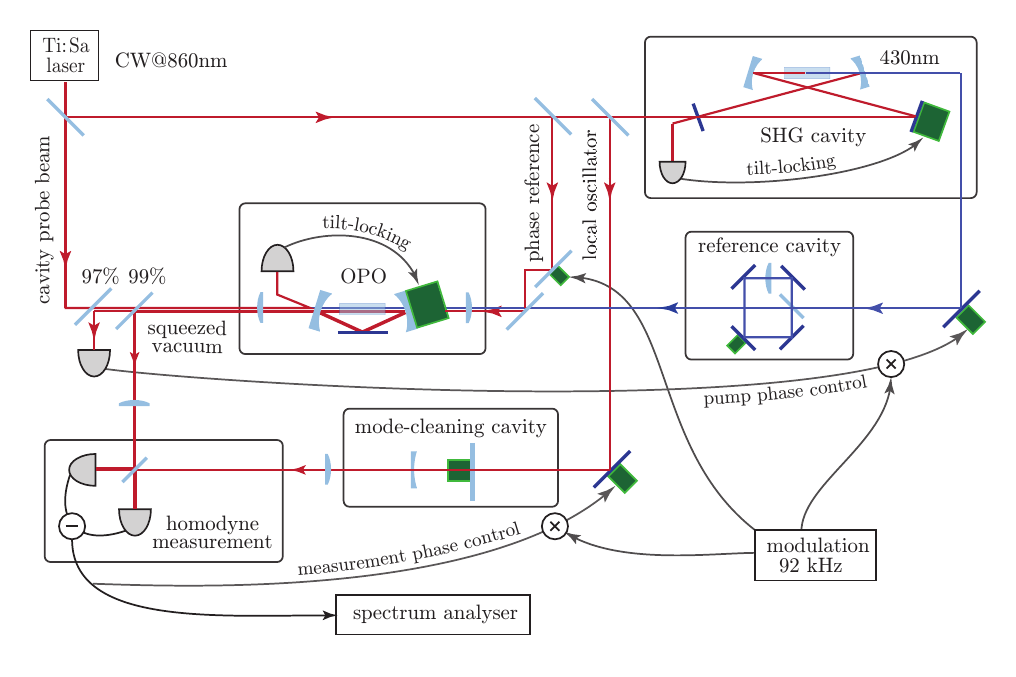}
 \caption{Schematic of the experimental setup. Red lines are the 860\,nm CW laser beams, and blue lines are the 430\,nm frequency-doubled beams. Only important optical elements such as beam splitters and mode-matching lenses are shown. Black lines denote electrical channels for measurement and control.} \label{fig:setup}
\end{figure}

The schematic of the experimental setup is shown in \fig\ref{fig:setup}. CW light at 860\,nm is produced by a Ti:Sapphire laser (M-squared, Sols:TiS). A bowtie-shaped second harmonic generator (SHG) supplies CW pump beam at 430nm with a type-I phase matched $\mathrm{KNbO}_3$ crystal. To match the transverse mode of the pump beam to the OPO, an auxiliary cavity is placed between the OPO and the SHG cavities as a reference of the beam shape. To generate a second harmonic light for alignment, a 860\,nm beam is temporally introduced in the opposite direction from the squeezed vacuum. The reference cavity is mode-matched to the second harmonic light of the OPO, and then the 430\,nm pump beam is adjusted to resonate to the reference cavity. During the measurement, the beam path of the reference cavity is blocked to prevent an unintended resonance.

The temperature of the PPKTP crystal is kept at $40^\circ \mathrm{C}$ for type-0 phase-matching. The OPO is continuously pumped to generate a squeezed vacuum through a sub-threshold degenerate parametric process. The squeezing and anti-squeezing spectra are obtained in balanced homodyne detection. A local oscillator (LO) beam is prepared as a circular $\mathrm{TEM}_{00}$ mode by a Fabry-Perrot type mode-cleaning cavity, providing a visibility of 99.1\% in the homodyne detection.

\begin{figure}
 \centering
 \includegraphics[width=6cm]{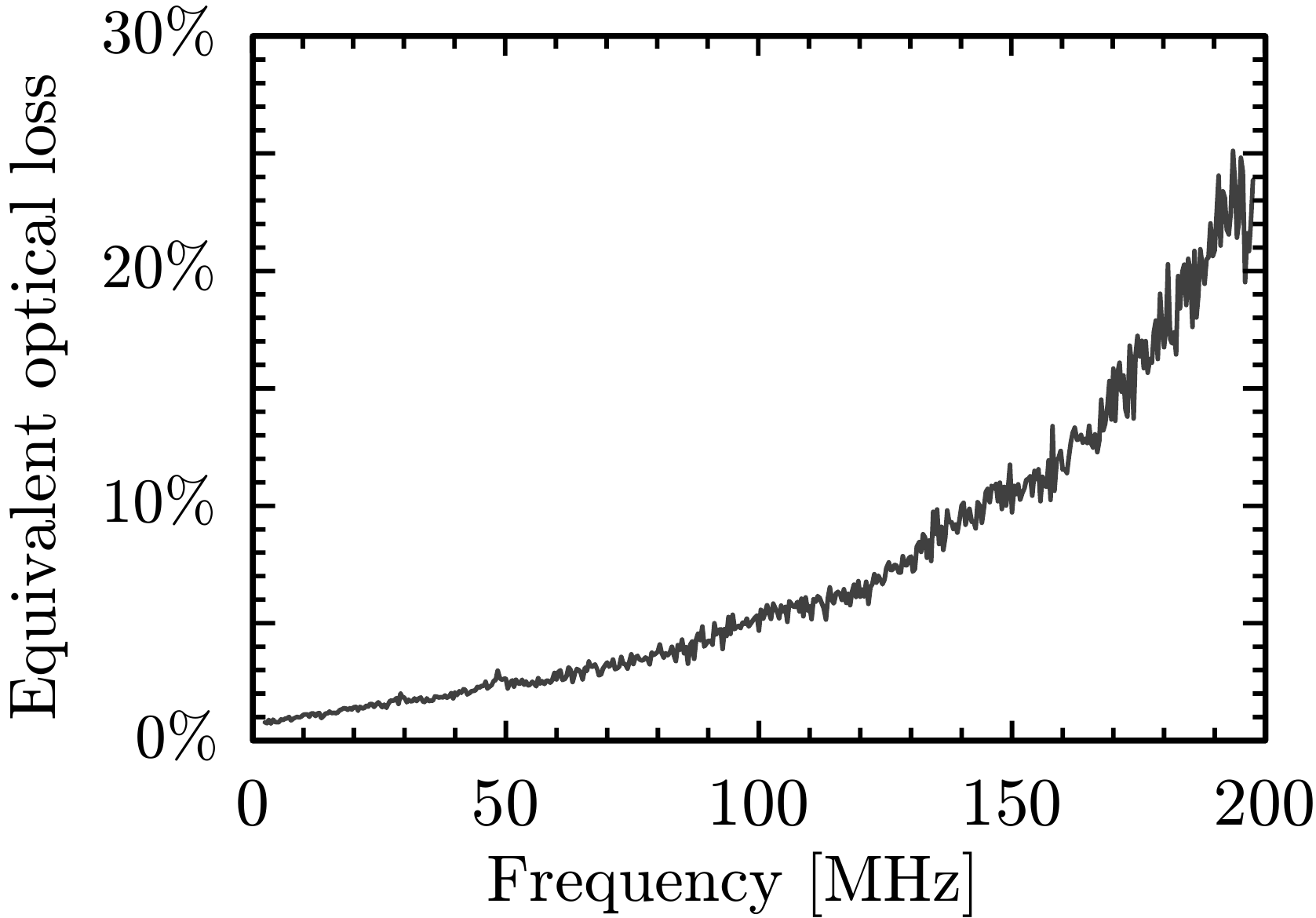}
 \caption{Equivalent optical loss of the electronical noise in the homodyne measurement. The local oscillator power is set at 18\,mW.} \label{fig:detloss}
\end{figure}

Low-loss, low-noise, and broadband homodyne detection is technically challenging. We make use of a wideband, flat-gain homodyne detector based on \cite{Lvovsky12}. A specially ordered photodiode (Hamamatsu Photonics, S5971SPL), which is anti-reflection coated at 860\,nm and whose quantum efficiency is higher than 98\%, is used. A 100\,V bias voltage is applied on it to shorten the drift time of carrier electrons, leading to the high-speed response above 100\,MHz. The active aperture diameter of 1.2\,mm is large enough to efficiently receive the signal light, while the high bias voltage minimizes the terminal capacitance to 1.5\,pF. 20\,dB of signal-to-noise ratio (SNR) is achieved in the shot noise detection near DC when the LO beam power is set at 18\,mW. \Figure\ref{fig:detloss} shows the equivalent optical loss spectrum of the electronical noise \cite{Appel07} calculated from the SNR spectrum obtained in the homodyne detection.

The OPO, the SHG and the mode-cleaning cavities are locked by tilt-locking technique \cite{shaddock99}, which requires no modulated probe field and consequently produces no undesired modulation signal in the squeezed vacuum beam. A phase-reference beam is introduced into the OPO along the pump beam, suffering parametric amplification according to the relative phase to the pump. 1\% of the reference beam (and also the squeezed light) is extracted through a partial-reflection mirror for monitoring, while the 1\% transmission allows a cavity-locking probe beam to run in the counter-propagating path to the squeezed vacuum. The phase signal of the pump beam appears as a 92\,kHz amplitude modulation in the reference beam, which originates from the phase modulation applied before getting in the OPO. After demodulating this signal, an error signal is obtained and fed back to the pumping phase, which means that the relative phase of the squeezed vacuum and the phase-reference beam is locked in parallel or orthogonal, depending on the polarity of error signal. The phase reference beam is also detected in the homodyne detection and provides the phase signal of the LO. A 300\,kHz high-pass filter is inserted before the spectrum analyzer to prevent saturation caused by the 92\,kHz modulation signal. All of these feedback controls are continuously performed during the measurement.

\section{Result}

\begin{figure}
 \centering
 \includegraphics[scale=1.1]{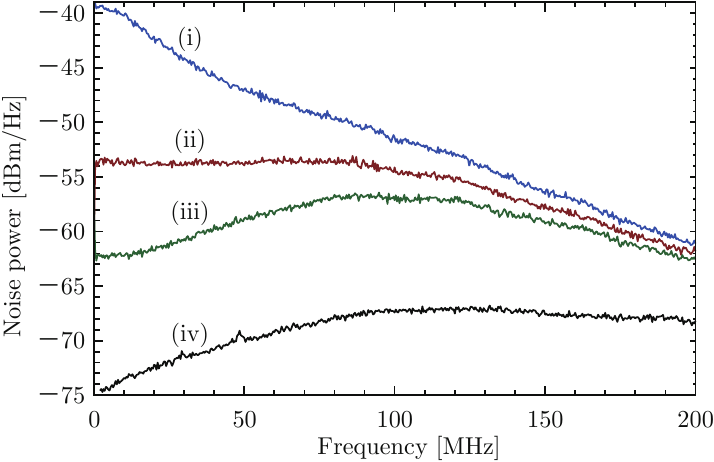}
 \caption{Quantum noise spectrum from the balanced homodyne measurement. (i) Anti-squeezing noise. (ii) Vacuum noise. (iii) Squeezing noise. (iv) Electronic noise (no LO light). This is a raw data from the spectrum analyzer, without any noise compensation. These data are taken with the resolution bandwidth of 300\,kHz, the video bandwidth of 300\,kHz, and averaged over 600 sweeps. The LO power is set at 18\,mW. DC--300\,kHz component is cut off by a high-pass filter to eliminate the reference beam modulation, so the low frequency component below 1MHz is not reliable.} \label{fig:spec}
\end{figure}

\begin{figure}
 \centering
 \includegraphics[scale=0.64]{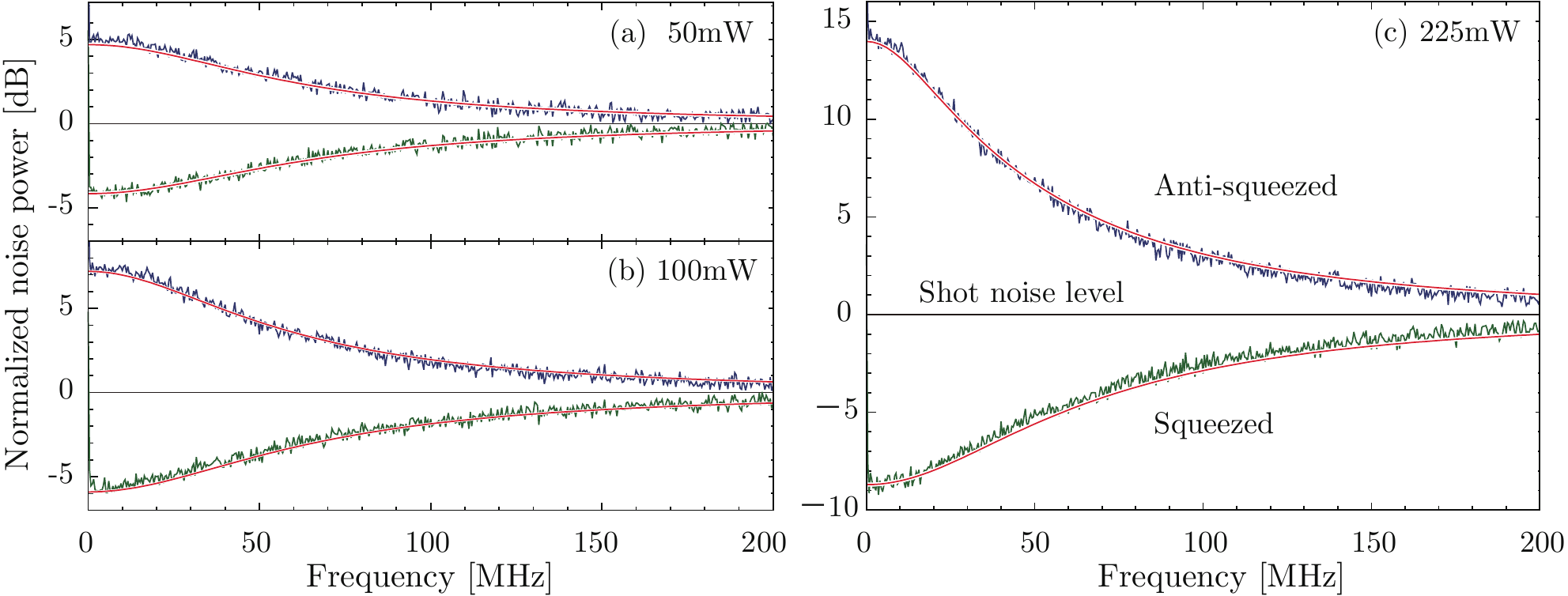}
 \caption{Normalized quantum noise spectrum for each pump powers. (a) Pump power is set at (a) 50\,mW, (b) 100\,mW, (c) 225\,mW. Red lines are theoretical prediction from \eq\eqref{eq:nspec}, \eqref{eq:phasefluc}, taking the detection losses and the phase fluctuation into account. Acquisition conditions are the same as \fig\ref{fig:spec}.} \label{fig:nspec}
\end{figure}

\Figure\ref{fig:spec} represents the quadrature variance spectrum of the vacuum field, the squeezed and anti-squeezed vacuum at the pump power of 225\,mW. $8.4$\,dB of squeezing around DC and $2.5$\,dB of squeezing at 100\,MHz is observed.
The normalized spectrums are shown in \fig\ref{fig:nspec}. The squeezing spectrum is well matched to the theoretical prediction, which is expressed as \cite{Collett84, Lam99}:
\begin{equation}
 R_\pm(f) = 1 \pm \eta \rho \frac{4\xi}{(1\pm\xi)^2+(f/f_\textrm{HWHM})^2} \label{eq:nspec}
\end{equation}
where $R_\pm$ is the anti-squeezing or squeezing level of the output from OPO, $f$ is the sideband frequency, $\eta$ is the total detection efficiency, $\rho = T / (T+L)$ is the escape efficiency of the OPO, $L$ is the internal cavity loss, and $\xi$ is the pump amplitude normalized by the oscillation threshold. The total detection efficiency $\eta$ depends on the frequency when we consider the electronic noise as an optical loss as shown in \fig\ref{fig:detloss}. Around DC, $\eta$ is 91.8\%, including 3.4\% propagation loss, 1.8\% mode-mismatch in the homodyne measurement, 2.0\% detection loss at photodiodes, and 1.0\% equivalent optical loss of the electronic noise. Escape efficiency $\rho$ is 98\%, derived from the internal cavity loss of 0.30\%.

\begin{figure}
 \centering
 \includegraphics[scale=0.55]{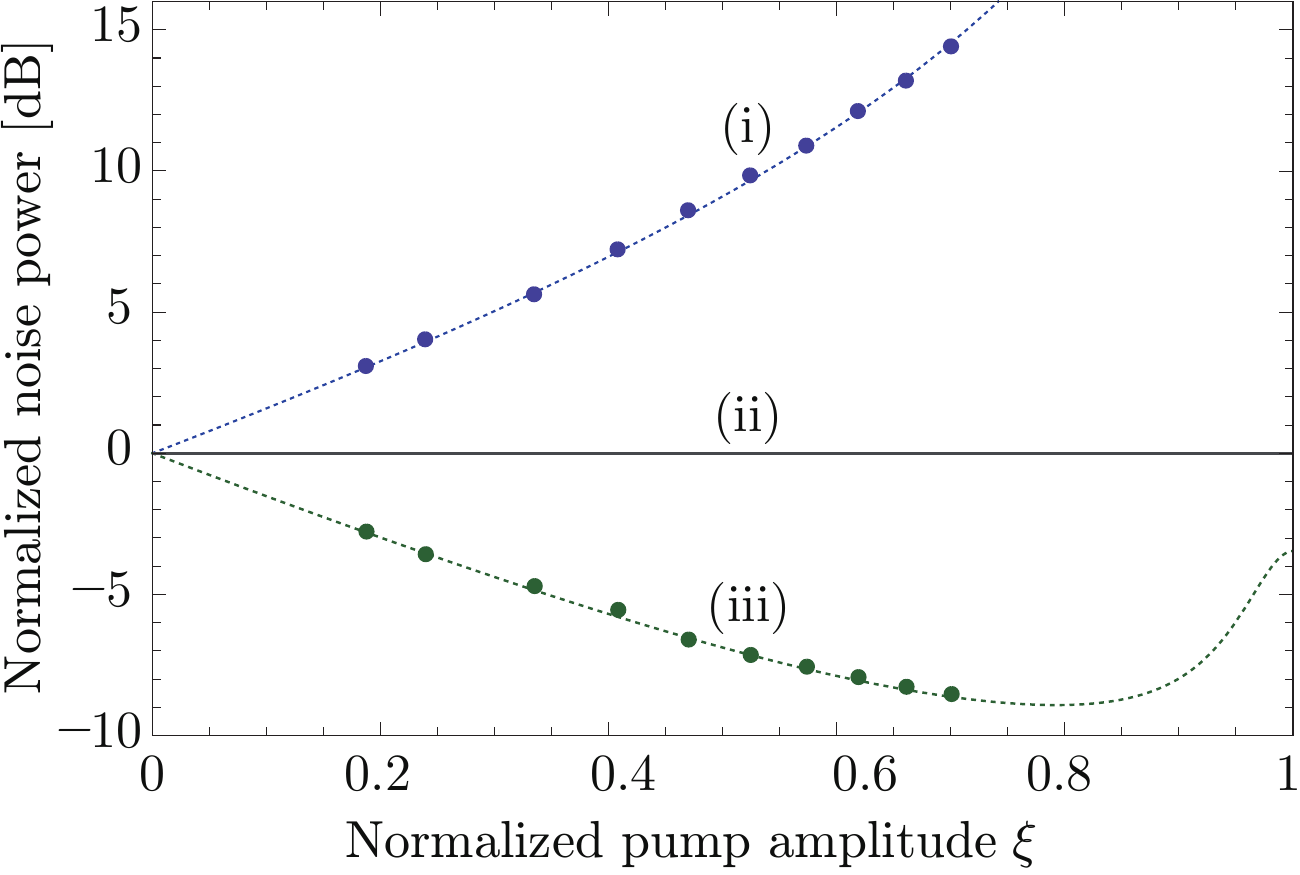}
 \caption{Pump amplitude dependence of the anti-squeezing and the squeezing level. (i) Anti-squeezing level. (ii) Shot noise level. (iii) Squeezing level. Dots denote the measured value at 3\,MHz. Dashed lines are theoretical prediction. The acquisition condition is as follows: resolution bandwidth is 30\,kHz, video bandwidth is 30\,kHz, and average number is 24,000.} \label{fig:bluepower}
\end{figure}
The pump power dependence of the squeezing level at 3\,MHz is shown in \fig\ref{fig:bluepower}. $8.4\,\mathrm{dB}$ of squeezing is realized at the pump power of 225\,mW. In this figure, the pump amplitude is normalized by the oscillation threshold, which is estimated at 490\,mW. Theoretical curve in this figure is calculated by \eq\eqref{eq:nspec}, considering the phase fluctuation as \cite{Takeno07}:
\begin{equation}
 R'_\pm(f) \sim R_\pm(f) \cos^2 \bar{\theta} + R_\mp(f) \sin^2 \bar{\theta} \label{eq:phasefluc}
\end{equation}
where $R'_\pm$ is squeezing and anti-squeezing level at homodyne detection, and $\bar{\theta}$ is the root-mean-square value of the phase fluctuation of the squeezed vacuum. We estimate $\bar{\theta}$ at $0.8^\circ$ from the phase monitor signals of the pump and the LO. The phase fluctuation do not have much effect on the squeezing level in this measurement because the pump power is kept under 225\,mW to prevent the PPKTP crystal from damage. However, the squeezing level will be limited to about $9$\,dB even under a higher pump power due to the phase fluctuation and the detection loss.

\section{Conclusion}
We develop a small, triangle-shaped ring OPO with 45\,mm round-trip length, which can generate 65\,MHz bandwidth squeezed vacuum. With a broadband homodyne detector, $8.4\,\mathrm{dB}$ of squeezing at 3\,MHz and $2.4$\,dB of squeezing at 100\,MHz is obtained without any noise compensation. The ring cavity structure of our OPO is suitable to be implemented in the large scale quantum optics setup with many OPOs. Thus, our OPO will contribute to the future high-speed quantum information processing schemes especially with the time-domain multiplexing technique.

\section*{Funding}
Japan Science and Technology Agency (JST) (CREST);
Japan Society for the Promotion of Science (JSPS) (KAKENHI JP26247066, KAKENHI JP16J09108);
Ministry of Education, Culture, Sports, Science and Technology (MEXT) (APSA, ALPS Program for Leading Graduate Schools);


\begin{thebibliography}{99}
\bibitem{Walls83} D. F. Walls, ``Squeezed states of light,'' \nat {\bf 306}(5939), 141--146 (1983).
\bibitem{Kimble01} H. J. Kimble, Y. Levin, A. B. Matsko, K. S. Thorne, and S. P. Vyatchanin, ``Conversion of conventional gravitational-wave interferometers into quantum nondemolition interferometers by modifying their input and/or output optics,'' \prd {\bf 65}(2), 022002 (2001).
\bibitem{Goda08} K. Goda, O. Miyakawa, E. E. Mikhailov, S. Saraf, R. Adhikari, K. McKenzie, R. Ward, S. Vass, A. J. Weinstein, and N. Mavalvala, ``A quantum-enhanced prototype gravitational-wave detector,'' {Nat.\ Phys.} {\bf 4}(6), 472--476 (2008).
\bibitem{Vahlbruch10} H. Vahlbruch, A. Khalaidovski, N. Lastzka, C. Gr\"{a}f, K. Danzmann, and R. Schnabel, ``The GEO 600 squeezed light source,'' {Class.\ Quantum\ Grav.} {\bf 27}(8), 084027 (2010).
\bibitem{Kolobov93} M. I. Kolobov, and P. Kumar, ``Sub-shot-noise microscopy: imaging of faint phase objects with squeezed light,'' \ol {\bf 18}(11), 849--851 (1993).
\bibitem{Lu12} Baozhu Lu, Siwen Bi, Fei Feng, Menghua Kang, and Fei Qin, ``Experimental study on the imaging of the squeezed-state light with a virtual object,'' {Opt.\ Eng.} {\bf 51}(11), 119001 (2012).
\bibitem{Ourjoumtsev06} A. Ourjoumtsev, R. Tualle-Brouri, J. Laurat, and P. Grangier, ``Generating Optical Schr\"{o}dinger Kittens for Quantum Information Processing,'' Science {\bf 312}(5770), 83--86 (2006).
\bibitem{Nielsen06} J. S. Neergaard-Nielsen, B. M. Nielsen, C. Hettich, K. M{\o}lmer, and E. S. Polzik, ``Generation of a Superposition of Odd Photon Number States for Quantum Information Networks,'' \prl {\bf 97}(8), 083604 (2006).
\bibitem{Wakui07} K. Wakui, H. Takahashi, A. Furusawa, and M. Sasaki, ``Photon subtracted squeezed states generated with periodically poled $\mathrm{KTiOPO_4}$,'' \opex {\bf 15}(6), 3568--3574 (2007).
\bibitem{Furusawa98} A. Furusawa, J. L. S{\o}rensen, S. L. Braunstein, C. A. Fuchs, H. J. Kimble, and E. S. Polzik, ``Unconditional Quantum Teleportation,'' Science {\bf 282}(5389), 706--709 (1998).
\bibitem{Lee11} N. Lee, H. Benichi, Y. Takeno, S. Takeda, J. Webb, E. Huntington, and A. Furusawa, ``Teleportation of Nonclassical Wave Packets of Light,'' Science {\bf 332}(6027), 330--333 (2011).
\bibitem{Yokoyama13} S. Yokoyama, R. Ukai, S. C. Armstrong, C. Sornphiphatphong, T. Kaji, S. Suzuki, J. I. Yoshikawa, H. Yonezawa, N. C. Menicucci, and Akira Furusawa, ``Ultra-large-scale continuous-variable cluster states multiplexed in the time domain,'' {Nat.\ Photonics} {\bf 7}(12), 982--986 (2013).
\bibitem{Yoshikawa16} Jun-ichi Yoshikawa, Shota Yokoyama, Toshiyuki Kaji, Chanond Sornphiphatphong, Yu Shiozawa, Kenzo Makino, and Akira Furusawa, ``Invited Article: Generation of one-million-mode continuous-variable cluster state by unlimited time-domain multiplexing,'' APL Photonics {\bf 1}(6), 060801 (2016).
\bibitem{Takeda12} Shuntaro Takeda, Takahiro Mizuta, Maria Fuwa, Jun-ichi Yoshikawa, Hidehiro Yonezawa, and Akira Furusawa, ``Generation and eight-port homodyne characterization of time-bin qubits for continuous-variable quantum information processing,'' \pra {\bf 87}(4), 043803 (2013).
\bibitem{Raussendorf01} Robert Raussendorf, and Hans J. Briegel, ``A One-Way Quantum Computer,'' \prl {\bf 86}(22), 5188--5191 (2001).
\bibitem{Gu09} Mile Gu, Christian Weedbrook, Nicolas C. Menicucci, Timothy C. Ralph, and Peter van Loock, ``Quantum computing with continuous-variable clusters,'' \pra {\bf 79}(6), 062318 (2009).
\bibitem{Takei06} Nobuyuki Takei, Noriyuki Lee, Daiki Moriyama, J. S. Neergaard-Nielsen, and Akira Furusawa, ``Time-gated Einstein-Podolsky-Rosen correlation,'' \pra {\bf 74}(6), 060101 (2006).
\bibitem{Slusher85} R. E. Slusher, L. W. Hollberg, B. Yurke, J. C. Mertz, and J. F. Valley, ``Observation of Squeezed States Generated by Four-Wave Mixing in an Optical Cavity,'' \prl {\bf 55}(22), 2409--2412 (1985).
\bibitem{Wu86} Liang-An Wu, H. J. Kimble, J. L. Hall, and Huifa Wu, ``Generation of Squeezed States by Parametric Down Conversion,'' \prl {\bf 57}(20), 2520--2523 (1986).
\bibitem{Yurke84} Bernard Yurke, ``Use of cavities in squeezed-state generation,'' \pra {\bf 29}(1), 408--410 (1984).
\bibitem{Collett84} M. J. Collett and C. W. Gardiner, ``Squeezing of intracavity and traveling-wave light fields produced in parametric amplification,'' \pra {\bf 30}(3), 1386--1391 (1984).
\bibitem{Anderson95} M. E. Anderson, M. Beck, M. G. Raymer, and J. D. Bierlein, ``Quadrature squeezing with ultrashort pulses in nonlinear-optical waveguides,'' \ol {\bf 20}(6), 620--622 (1995).
\bibitem{Serkland95} D. K. Serkland, M. M. Fejer, R. L. Byer, and Y. Yamamoto, ``Squeezing in a quasi-phase-matched LiNbO3 waveguide,'' \ol {\bf 20}(15), 1649--1651 (1995).
\bibitem{Eto11} Y. Eto, A. Koshio, A. Ohshiro, J. Sakurai, K. Horie, T. Hirano, and M. Sasaki, ``Efficient homodyne measurement of picosecond squeezed pulses with pulse shaping technique,'' \ol {\bf 36}(23), 4653--4655 (2011).
\bibitem{Shelby86} R. M. Shelby, M. D. Levenson, S. H. Perlmutter, R. G. DeVoe, and D. F. Walls, ``Broad-Band Parametric Deamplification of Quantum Noise in an Optical Fiber,'' \prl {\bf 57}(6), 691--694 (1986).
\bibitem{Bergman91} K. Bergman, and H. A. Haus, ``Squeezing in fibers with optical pulses,'' \ol {\bf 16}(9), 663--665 (1991).
\bibitem{Vahlbruch16} Henning Vahlbruch, Moritz Mehmet, Karsten Danzmann, and Roman Schnabel, ``Detection of 15 dB Squeezed States of Light and their Application for the Absolute Calibration of Photoelectric Quantum Efficiency,'' \prl {\bf 117}(11), 110801 (2016).
\bibitem{Breitenbach95} G. Breitenbach, T. M\"{u}ller, S. F. Pereira, J.-Ph. Poizat, S. Schiller, and J. Mlynek, ``Squeezed vacuum from a monolithic optical parametric oscillator,'' \josab {\bf 12}(11), 2304--2309 (1995).
\bibitem{Ast13} S. Ast, M. Mehmet, and R. Schnabel, ``High-bandwidth squeezed light at 1550\,nm from a compact monolithic PPKTP cavity,'' \opex {\bf 21}(11), 13572--13579 (2013).
\bibitem{Laurat05} J. Laurat, T. Coudreau, and C. Fabre, ``Type-II Optical Parametric Oscillator : a versatile source of quantum correlations and entanglement,'' https://arxiv.org/abs/quant-ph/0510063.
\bibitem{Zhou15} Yaoyao Zhou, Xiaojun Jia, Fang Li, Changde Xie, and Kunchi Peng, "Experimental generation of 8.4 dB entangled state with an optical cavity involving a wedged type-II nonlinear crystal," \opex {\bf 23}(4), 4952-4959 (2015).
\bibitem{Takeno07} Y. Takeno, M. Yukawa, H. Yonezawa, and A. Furusawa, ``Observation of $-9$\,dB quadrature squeezing with improvement of phase stability in homodyne measuremnt,'' \opex {\bf 15}(7), 4321--4327 (2007).
\bibitem{Lvovsky12} R. Kumar, E. Barrios, A. MacRae, E. Cairns, E. H. Huntington, and A. I. Lvovsky, ``Versatile wideband balanced detector for quantum optical homodyne tomography,'' \oc {\bf 285}(24), 5259--5267 (2012).
\bibitem{Appel07} J. Appel, D. Hoffman, E. Figueroa, and A. I. Lvovsky, ``Electronic noise in optical homodyne tomography,'' \pra {\bf 75}(3), 035802 (2007).
\bibitem{shaddock99} D. A. Shaddock, M. B. Gray, and D. E. McClelland, ``Frequency locking a laser to an optical cavity by use of spatial mode interference,'' \ol {\bf 24}(21), 1499--1501 (1999).
\bibitem{Lam99} P. K. Lam, T. C. Ralph, B. C. Buchler, D. E. McClelland, H. A Bachor, and J. Gao, ``Optimization and transfer of vacuum squeezing from an optical parametric oscillator,'' J. Opt. B {\bf 1}(4), 469--474 (1999).
\end{thebibliography}
\end{document}